# How Quantum Theory Helps us Explain

Richard Healey

University of Arizona,

213 Social Sciences,

Tucson,

AZ 85721-0027,

USA.

**Email:** rhealey@email.arizona.edu

**Phone:** +1 520-975-7053

**Fax:**   +1 520-621-9559

# How Quantum Theory Helps us Explain

'In the beginning natural philosophers tried to understand the world around them. … Experimental science was born. But experiment is a tool. The aim remains: to understand the world. To restrict quantum mechanics to be exclusively about piddling laboratory operations is to betray the great enterprise.'

        J.S. Bell, 'Against "measurement" ' ([2004], p. 217).

## 1. Introduction

While the great predictive power of quantum theory is universally acknowledged, its explanatory credentials are still actively debated among those concerned with the theory's conceptual foundations. There are instrumentalists who restrict the scope of quantum theory to a set of rules for calculating probabilities for the outcomes of (macroscopic) tests which follow specified (macroscopic) preparations.[1] Bell argued forcefully against such instrumentalists in the paper quoted above as well as several other influential pieces collected in his [2004]. In his view, only if reformulated precisely in terms of a clear ontology of "beables" could quantum theory supply the kind of explanations we need to understand the big world outside the laboratory. Others attempting to portray quantum theory as offering fundamental explanations as well as descriptions of reality have been driven to champion an Everettian interpretation of the theory, Bohmian mechanics or objective collapse theories.[2] The continuing failure to agree on any specific realist interpretation or reformulation contrasts strikingly with the widespread acceptance in the scientific community of the enormous *explanatory* power of contemporary quantum theory. This has prompted some to seek an interpretation-neutral account of explanation in quantum theory in structural terms.[3] But they must face Bell's objection that, unless reformulated as a precise theory about beables, quantum theory is merely a blunt instrument for predicting statistics of measurement results obtained in ill-defined circumstances that is consequently unable to explain anything that happens outside the laboratory.

The pragmatist view of quantum theory I have outlined elsewhere (Healey [2012a]) presents a middle way between realism and instrumentalism with the resources to answer Bell's charge and show how quantum theory advances the great aim of understanding the world. Viewed this way, quantum theory helps us explain a huge variety of natural phenomena in and outside the laboratory. A distinctive feature of such explanations is that quantum theory does not itself extend our resources for representing or describing the systems these phenomena involve. We are able to use quantum models to show that otherwise surprising phenomena were to be expected and to say what they depend on. By repeatedly using the same model elements and similar kinds of models we have been able to unify our understanding of otherwise diverse phenomena. Perhaps this makes it appropriate to call explanations using quantum theory structural, even if quantum theory does not itself represent novel structures underlying the phenomena it helps us to explain.

This paper proceeds as follows. Section 2 suggests two general requirements on explanations in physics while noting that there are different ways of meeting them. Section 3 lists a small but representative sample of the extraordinary variety of phenomena that are generally thought to be explained by quantum theory. In section 4 I review the main points of the pragmatist interpretation of quantum theory outlined in (Healey [2012a]), highlighting what it takes to be the functions of the quantum state and the Born probability rule. Section 5 applies this pragmatist interpretation to give a general account of how quantum theory can help explain otherwise puzzling regularities and shows how far such explanations meet the two requirements noted in section 2. In each of the next three sections this account is illustrated by applying it to a notable explanatory success of quantum theory: single particle interference phenomena in section 6, the stability of matter in section 7, and Bose-Einstein condensation in section 8. In conclusion I note some open problems and relate the present account of explanation in quantum theory to alternative approaches that emphasize the importance of causation, unification, and structure.

## 2. Two requirements on explanations in physics

Explanations in physics may turn out to have no more in common than games, the example Wittgenstein used to introduce his notion of family resemblance. The analogy between scientific explanations and family members extends further. Resemblances within a family change through the generations. New members join the family by birth, marriage or adoption and old members die or are otherwise estranged. So, too, what we take to be a satisfactory explanation has changed during the development of physics, and we may confidently expect such change to continue. According to the pragmatist interpretation of (Healey [2012a]), one who accepts quantum theory is able to offer a novel kind of explanation. Nevertheless, explanations of phenomena using quantum theory may be seen to meet two very general conditions met by many, if not all, good explanations in physics. (i) They show that the phenomenon to be explained was to be expected and (ii) they say what it depends on.

What does it mean to say that the occurrence of the phenomenon *was to be expected*—expected by whom, given what? A proponent of the DN account of scientific explanation could answer that the phenomenon was to be expected by anyone knowing the relevant laws and specifications of particular circumstances and able to reason from them. But critics have objected that such an argument is genuinely explanatory only if its premises state what caused the phenomenon, where it is assumed that any cause of an instance of the phenomenon bears an asymmetric relation of causal influence to it. Lewis ([1986]) argued that to explain a particular event is to provide information about its causal history: the event would be expected by anyone who had enough such information ahead of time. A general regularity constituted by events of certain kinds could have been anticipated by an agent armed with sufficient information about the causal history of events of these kinds.[4] To provide such information would be to say what the regularity (causally) depends on.

Woodward ([2003]) offered an influential interventionist analysis of causal explanation that leaves room for non-causal scientific explanations of regularities such as a possible explanation of the stability of planetary orbits by appeal to the dimensionality of space-time. He proposed a more general requirement on a scientific explanation that is capable of accommodating this and other non-causal explanations of regularities[5]: a scientific explanation should provide the information needed to answer what he calls *w-questions*.

> the common element in many forms of explanation, both causal and non-causal, is that
> 
> they must answer what-if-things-had-been-different questions. (*op. cit.*, p. 219)

Things might have been different if the events figuring in the causal history of a general regularity had been different in various ways. But they would also have been different if energy had not been conserved, if parity *had* been conserved, or (perhaps) if there had been no magnetic monopoles. To answer a *w*-question one may toy with events in the causal history of a phenomenon. Or one may consider alternatives to some general law, fact or regularity.

An answer to a *w*-question about a general regularity says what it (counterfactually) depends on, and how it depends on it: it embeds the phenomenon in a pattern of functional dependencies,และthereby explains it. The regularity was to be expected by one who knows this pattern and knows the values of the variables in the pattern on which it depends. An explanation of a regularity is causal only if at least some of these counterfactuals expressing dependency relations may be understood causally.[6]

Assuming the pragmatist interpretation sketched in section 4, section 5 shows how quantum theory helps us to see that many puzzling phenomena were to be expected and enables us to say what each phenomenon depends on and so to answer a variety of *w*-questions about it. This will exhibit novel senses of expectation and dependence as compared to theoretical explanations provided by classical physics. To better appreciate the novelty, it will be useful to appeal to what Hughes ([2010]) presents as a general account of theoretical explanation that fits classical physics, but not, I shall argue, quantum theory.

Hughes ([2010], p. 210) offers this encapsulation of his general account of theoretical explanation.

> We explain some feature *X* of the world by displaying a theoretical model *M* of part of the world, and demonstrating that there is a feature *Y* of the model that corresponds to *X* and is not explicit in the definition of *M*.

Hughes's account correctly emphasizes the pivotal role played by models in theoretical explanations within physics, treating laws as referring to model elements and having no theory-independent interpretation. And it nicely captures an intuition that

> ...theoretical explanations allow us to understand the world, not by showing its conformity to principles external to the theory, but by representing it in terms of the model the theory itself supplies. As we become aware of the resources of these representations, so we come to understand the phenomena they represent. Hence ... the greater the variety of contexts in which we see the theory applied, the deeper this understanding becomes. (p. 231)

This is how 'it endorses our intuition that theoretical unification brings about an increase in understanding.' (p. 226)

Notice the key role representation plays here: feature $Y$ of the model corresponds to $X$ *by representing it*. Hughes even calls his accompanying version of the so-called semantic conception of scientific theories the representational account. A Newtonian model aids understanding by allowing us to think of parts of the solar system, their masses and motions and the forces between them, in terms of the individual element of the model by which each is represented. This at least permits, and perhaps encourages, a realist attitude toward structures newly introduced by the theory, such as masses and gravitational forces. For while representation does not require the reality of that which is represented, it at least provides a launching pad for an inference to the existence of what is represented in the best explanation of the phenomena.

The use of Newton's theory of motion and gravitation to explain the regular motions of the planets is a paradigm case of how theoretical explanation functions in classical physics. Kepler's three laws are the regularities to be explained. These are initially specified as regularities in the spatiotemporal trajectories of astronomical bodies, with reference neither to their constitution and masses nor to the forces acting on them. First one uses the terminology of Newton's theory to *re*present a planet as composed of a vast number of massive particles, each acted on by the gravitational force exerted by all the particles composing a much more massive sun. Then one constructs a class of idealized mathematical models of a fixed sun and a single planet in conformity to Newton's laws of motion and gravitation. One demonstrates mathematical relations holding in these models that one interprets as corresponding to Kepler's laws. One notes that while one could not demonstrate exactly these mathematical relations within less idealized models denoting a movable sun with other planets and objects in the solar system, there is reason to believe that approximations to these relations hold also in such models. One concludes that Kepler's laws hold in the solar system *to a good approximation*.

The model demonstration may be considered to exhibit the dependence of Kepler's laws on Newton's laws: Kepler's laws would not have held if force had equaled mass x (acceleration)$^2$, or gravity had obeyed an inverse cube law or not been a central force. This is asymmetric insofar as (unlike Kepler's laws) Newton's laws of motion and gravitation are common to a great variety of models otherwise unrelated to these planetary models, thereby helping us explain many different regularities. The demonstration also shows how Kepler's laws depend on the particular circumstances obtaining in the solar system. It is readily extended to show that the orbit of a planet that moved much faster would not approximate an ellipse but a parabola or hyperbola: and that there would be no reason to expect Kepler's laws to hold if unsuspected non-gravitational forces were present in the solar system.

The demonstration also shows the sense in which Kepler's laws were to be expected. They were to be expected by anyone knowing the structure of models enshrining Newton's laws of motion and gravitation; the material nature, approximate relative masses and momentary states of the sun and planets; and the absence of non-gravitational planetary forces in the solar system.

In this and other examples of theoretical explanation using classical physics, one begins with a claim describing or otherwise representing the regularity to be explained together with some (implicit if not explicit) specification of the circumstances in which it obtains. The explanation of the regularity is provided by reference to one or more models provided by the explanatory theory. Such models not only *re*present the regularity in terms of the explanatory theory, but introduce additional theoretical elements that may be taken to represent aspects of the world that underlie the regularity. That is how Newton explained (the approximate validity of) Kepler's laws by appeal to theoretical models that included not only mathematical representations of the planetary trajectories whose features were to be explained, but also mass and gravitational force parameters representing additional basic magnitudes.

To use a physical theory, whether classical or quantum, to explain a regularity one must describe or otherwise represent systems that manifest the regularity as well as the circumstances in which it obtains: but that representation need not be novel to the theory one uses to explain it. Indeed, to be acknowledged as available for potential explanation the regularity *must* be specifiable independently of the theory that is to be used to explain it. This prior representation may be provided by some other theory, or in a language or representational system not associated with any recognized theory. It is common in physics to "nest" successive representations of the phenomenon to be explained, sometimes starting with a rough description such as 'high temperature superconductivity in iron pnictides', then representing the phenomenon within the framework of successive theories in progressively more abstract and idealized terms until one arrives at a representation to which a theory can be applied to (try to) explain the phenomenon.

For an explanation using *classical* physics, the explanatory theory then supplies one or more models that can be used to *re*present the regularity in explaining it. Objects, events and processes figuring in the regularity as well as their detailed features are denoted by corresponding elements of a theoretical model, many of them mathematical. Among other things, this theoretical representation will often improve the initial specification of the regularity. It may correct this, or enrich it by bringing out features that show the regularity's relation to other phenomena. But this is not essential: nor is it the only way to improve an initial representation[7].

So representation plays a twofold role in theoretical explanations within classical physics. Its primary role is realized by terms used to represent the phenomena to be explained. But theoretical models typically also include terms that (at least purport to) represent novel physical structures. The theoretical terms introduced in this secondary representational role then figure essentially in the demonstration within the model of the representation of the phenomenon that is thereby explained. As we shall see in section 5, according to the pragmatist interpretation of (Healey [2012a]) novel terms in quantum theoretic models do not play such a secondary representational role, and there is nothing quite like a demonstration within a quantum model of a representation of the phenomenon we use it to explain.

**3. What we can use quantum theory to explain**

The explanatory power of quantum theory is without parallel in the history of physics. From Schrödinger's explanation of the energy levels of the hydrogen atom, explanatory applications of quantum theory have now extended to systems as disparate as the energy levels of quark-antiquark systems, samples of superfluid $^3$helium, the interaction of electrons with light, ferromagnets, complex organic molecules, transistors, various kinds of quantum vacuum, lasers, neutron stars, mesoscopic mirrors, entangled photon pairs separated by many kilometers, and even the inflaton field whose fluctuations may have given rise to the large scale distribution of matter in the universe at galactic and super-galactic scales. Clearly it is impossible here to undertake a comprehensive survey of all these applications. Instead I will simply list a number of representative cases in which quantum theory has helped us explain otherwise puzzling phenomena.

Apart from their diversity, three aspects of these cases are worth emphasizing. First, each case itself splits up into many sub-cases, corresponding to the variety of features we can use quantum theory to explain here. Second, classical physics dramatically fails to offer any explanation of most of these phenomena; and even for those for which it could offer an explanation the quantum explanation is generally both superior and quite different. Note finally that while some of these phenomena have been observed only under carefully controlled laboratory conditions, there is no reason to doubt that others occur naturally, and some seem clearly beyond our powers to confine to any laboratory!

- The existence and detailed properties of interference displayed by single electrons, neutrons, photons, $C_{60}$ molecules, etc.

- Why ordinary atomic matter is stable.
- Bose-Einstein condensation, and the interference between certain separately prepared samples of a dilute gas BEC.
- The frequencies and relative intensities of spectral lines.
- The shape of the black body spectrum.
- The temperature-dependence of the specific heat of a solid.
- The structure of the periodic table.
- Features of the chemical bond, e.g. in molecular hydrogen and benzene.
- The broad division between conductors, insulators, semiconductors and superconductors, as well as detailed properties of each.
- The broad division between paramagnetic, diamagnetic, ferromagnetic, anti-ferromagnetic and ferrimagnetic materials, as well as detailed properties of each.
- The half-lives as well as other features of various kinds of radioactive decay processes.
- Superfluidity of various kinds, and the detailed properties of each.
- Laser action and details of laser operation.
- Properties of white dwarfs and neutron stars.
- Features of the charmonium and bottomonium spectra.
- The formation of tracks in a bubble chamber or spark chamber.
- Violations of Bell inequalities.
- The formation of structure in the very early universe.

- How quantum teleportation and secure public-key distribution are possible, and how a quantum computer *might* be able to execute certain algorithms faster than any feasible classical computer.

This list of explanatory successes is nothing if it is not impressive. But it may be nothing. Scientists as well as philosophers have periodically rejected the explanatory claims even of extremely successful physical theories. It is easy to ignore the obduracy of philosophers such as Duhem and (the early) Wittgenstein by attributing it to their unattainably high *a priori* requirements on explanation.[8] But, as historians have pointed out, radical theoretical developments have regularly prompted changes in standards of explanatory adequacy, resisted by some of the best physicists of their generation.[9] It is among those most concerned with the conceptual foundations of quantum theory that one is likely to meet the greatest resistance to the theory's explanatory claims. The goal of this paper is to marshal the resources of the pragmatist interpretation of (Healey [2012a]) to combat such explanatory skepticism.

## 4. The function of quantum states and Born probabilities

In this section I sketch the pragmatist interpretation of quantum theory outlined more fully in (Healey [2012a]) that will be assumed in the rest of the paper.

In quantum theory as it is usually formulated, theoretical models involve wave-functions (or more general mathematical objects) and operators corresponding to observables (including the Hamiltonian and/or Lagrangian) and (solutions to) the Schrödinger equation and relativistic generalizations. But there is still no agreement as to how, or whether, any of these model elements represent basic physical magnitudes. Einstein argued that the wave-function gives an incomplete description of physical reality, but von Neumann and others defended its descriptive completeness. Dirac's distinction between *q*-numbers and *c*-numbers merely labeled the problem posed by the difference in representational status of quantum observables and classical dynamical variables without solving the problem. If theoretical explanation requires representation in a model of some theory, an adequate account of quantum explanation must address these issues.

One radical position is that neither wave-functions nor operators nor dynamical laws represent or describe the condition or behavior of the physical system to which they pertain. The pragmatist interpretation of quantum theory I outline in (Healey [2012a]) adopts this radical position. It accepts the implication that quantum theory by itself explains nothing: the novel elements appearing in its models cannot be used to demonstrate representations of the phenomena to be explained. For quantum theory does not imply statements one can use to make claims about natural phenomena that describe or represent features of those phenomena. But quantum theory nevertheless helps us to explain an extraordinary variety of regularities in the physical world using representational resources from outside of quantum theory. It can do this because there is more to the informational structures quantum theory supplies than theoretical models involving wave-functions, Hamiltonian, Lagrangian and other operators, and (solutions to) the Schrödinger equation and relativistic generalizations. Implicit in the theory are rules for *using* these models to guide one in making claims and forming beliefs about physical systems to which such models may be applied *but which these models do not themselves describe or represent*.

The dispute as to whether quantum theory provides a complete description of a physical system presupposes that wave-functions at least provide a *partial* description or representation of the physical properties of systems to which they are assigned. Rejecting this presupposition may seem tantamount to adopting an instrumentalist interpretation of the wave-function or other mathematical representative of the quantum state—i.e. one that regards this as merely a symbolic device for calculating probabilities of possible measurement outcomes on these systems. But this is not so. Assignment of a quantum state may be viewed as merely the first step in a procedure that licenses a user of quantum theory to express claims about physical systems in descriptive language and then warrants that user in adopting appropriate epistemic attitudes toward these claims. The language in which such claims are expressed is not the language of quantum states or operators, and the claims are not about probabilities or measurement results: they are about the values of magnitudes. That is why I refer to such claims as NQMCs—Non-Quantum Magnitude Claims. Here are some typical examples of NQMCs[10]:

> A helium atom with energy $-24.6$ electron volts has zero angular momentum.
>
> Silver atoms emerging from a Stern-Gerlach device each have angular-momentum component either $+\hbar/2$ or $-\hbar/2$ in the $z$-direction.
>
> The fourth photon will strike the left-hand side of the screen.
>
> When a constant voltage $V$ is applied across a Josephson junction, an alternating current $I$ with frequency $2(e/h)V$ flows across the junction.

(Notice that two of these *non-quantum* claims are stated in terms of Planck's constant.) For contrast, here are some quantum mechanical claims which do *not* describe the physical properties of systems to which they pertain:

> The expectation value of angular momentum for atomic helium in the ground state is 0.
>
> Integral-spin systems have symmetric quantum states, while half-integral-spin systems have antisymmetric quantum states.
>
> The probability that a tritium nucleus will decay in 12.3 years is ½.
>
> After one photon from the polarization-entangled Bell state $|\Phi^+\rangle$ is found to be horizontally polarized, the other photon has polarization state $|H\rangle$.

This is not to say that quantum mechanical claims like these lack truth-values—each is appropriately evaluated as true (though in the case of the last claim that evaluation is critically dependent on the context relative to which it is made). But the *function* of such claims is not to describe or represent properties of physical systems: it is to offer authoritative advice to a physically situated agent on the content and credibility of NQMCs concerning them.[11] Quantum theory contributes indirectly to our explanatory projects.

Physicists and other humans are physically situated agents in a position to benefit from quantum advice in a wide variety of circumstances. They are at present the only users of quantum theory, individually or collectively. But nothing rules out the possibility of non-human, or even non-conscious, users of quantum theory, provided these are physically situated. This proviso is required because a quantum state and the consequent Born probabilities can be assigned to a system only relative to the physical situation of an (actual or hypothetical) agent for whom these assignments would yield good epistemic advice. What one agent should believe may be quite different from what another agent in a different physical, and *therefore* epistemic, situation should find credible. This relational character of quantum states and Born probabilities does not make these subjective, and it may be neglected whenever users of quantum theory find themselves in relevantly similar physical situations, which is often the case when we use quantum theory to help us explain a regularity, including those considered in sections 6-8. NQMCs are also objective, but, unlike claims pertaining to quantum states and Born probabilities, they are *not* relational: Their truth-values do not depend on the physical situation of any actual or hypothetical agent.

On the present pragmatist understanding, a quantum state guides an agent in two different ways. The agent requires guidance in assessing the content of NQMCs about systems of interest in a context where such claims may arise. It is often said that assignment of a value to an observable on a system is meaningful only in the presence of some apparatus capable of measuring the value of that observable. But some account of meaning must be offered in support of this assertion, and the extreme operationist account that is most naturally associated with it would be unacceptably vague even if it were otherwise defensible. Exactly what *counts* as the presence of an apparatus capable of measuring the value of an observable?

Contemporary pragmatist accounts of meaning have the resources to provide a better account of the meaning of a NQMC about a system, as entertained by an agent, in a context in which that system features. A pragmatist like Brandom ([1994], [2000]) takes the content of any claim to be articulated by the material inferences (practical as well as theoretical) in which it may figure as premise or conclusion. These inferences may vary with the context in which a claim arises, so the content of the claim depends on that context. The quantum state modulates the content of NQMCs about a system by specifying the context in which they arise. The context may be specified by the nature and degree of environmental decoherence suffered by this quantum state. A NQMC about a system when the quantum state has extensively decohered in a basis of eigenstates of the operator corresponding to that magnitude on the system has a correspondingly well-defined meaning: a rich content accrues to it *via* the large variety of material inferences that may legitimately be drawn to and from the NQMC in that context. I shall call a NQMC *canonical* if it is of the form

$$\text{Magnitude } M \text{ has value in Borel set } \Delta \text{ of real numbers}$$

and I shall write this as $M\varepsilon\Delta$. Only when the content of a canonical NQMC is sufficiently well articulated in this way is it appropriate to apply the Born Rule to assign a probability to that claim.

For example, a claim that an electron passed unobservably through a particular slit in a diffraction grating figures as premise or conclusion in almost no interesting material inferences, and so is very poorly articulated. This is a consequence of the fact that, in the absence of interactions capable of revealing its presence, the quantum state of the electron suffers negligible decoherence through entanglement with the environment in the interferometer. With no significant decoherence between different spatially localized quantum states, an agent has only a very limited license to use claims about the electron's position. In particular, the license is so limited that little may legitimately be inferred from such claims. But the subsequent interaction with detectors at the screen involves massive environmentally induced decoherence of the quantum state of the detector, permitting a high degree of articulation of the content of claims about an electron's position at the screen. The decohered quantum state then grants an agent a wide license to formulate and use such claims. This illustrates the first way in which a quantum state guides an agent—by advising her on the extent of her license to use particular claims by informing her of the nature and degree of that state's environmental decoherence.

With a sufficiently extended license, an agent may now apply the Born Rule to evaluate the probability of each licensed NQMC of the form $M\varepsilon\Delta$ using the appropriate quantum state. In the example, this will be the initial superposed state of the electron. Lacking more direct observational information, after assigning the quantum state of a system appropriate to her physical situation, an agent should adjust her credences in licensed NQMCs pertaining to that system so they match the probabilities of NQMCs specified by the Born Rule. This is the second way in which a quantum state guides an agent.

An agent should accept this twofold guidance by the right quantum state for one in her physical situation, since by doing so she is able successfully to predict and (as we shall see) explain what happens in a wide variety of circumstances. But according to the pragmatist interpretation of Healey ([2012a]) she need not, and should not, assume that this quantum state describes or represents novel intrinsic features of a system to which she ascribes it. The interpretation does not deny that a particular process occurs when a physical system interacts with its physical environment. But it does deny that the function of a model of decoherence of a system's quantum state is to represent what happens to that system in this process. An agent appeals to the quantum theory of decoherence to decide what to think about the content and credibility of NQMCs about a physical system, not to describe the evolution of a system whose quantum state decoheres.

Before going on to show how this pragmatist view of quantum theory helps us explain, let me clearly distinguish it from two better known alternatives: instrumentalism and constructive empiricism.

On the present pragmatist view, given the quantum state there may be some canonical NQMCs on a system to which the Born Rule is legitimately applied. Notice that neither a NQMC nor the Born Rule assigning it a probability here involves any explicit or implicit reference to measurement; nor does specification of the conditions under which the Born Rule may legitimately be applied. In this pragmatist view, the Born Rule does *not* apply only to results of measurements. This distinguishes the view from any Copenhagen-style instrumentalism according to which the Born Rule assigns probabilities only to possible outcomes of measurements, and so has nothing to say about unmeasured systems. An agent may use quantum theory to adjust her credences concerning what happened to the nucleus of an atom long ago on an uninhabited planet orbiting a star in a galaxy far away, provided only that she takes this to have happened in circumstances when that nucleus's quantum state suffered suitable environmental decoherence.

According to current usage, instrumentalism in the philosophy of science is the view that a theory is merely a tool for systematizing and predicting our observations: For the instrumentalist, nothing a theory supposedly says about unobservable structures lying behind but responsible for our observations should be considered significant. Moreover, instrumentalists characteristically explain this alleged lack of significance in semantic or epistemic terms: claims about unobservables are meaningless, reducible to statements about observables, eliminable from a theory without loss of content, false, or (at best) epistemically optional even for one who accepts the theory. The pragmatist view of Healey ([2012a]) makes no use of any distinction between observable and unobservable structures, so to call it instrumentalist conflicts with current usage.

In the pragmatist view of Healey ([2012a]), quantum theory does not posit novel, unobservable structures corresponding to quantum states, operators and Born probabilities: these are not physical structures at all.[12] Nevertheless, claims about them in quantum theory are often perfectly significant, and many (including the four earlier examples of quantum mechanical claims) are true. The pragmatist view of Healey ([2012a]) does not seek to undercut the semantic or epistemic status of such claims, but to enrich our understanding of their non-representational function within the theory and to show how they acquire the content they have.

There is a widespread view that the role of the wave-function (or more general mathematical object) is to represent a novel physical structure—the quantum state—whose existence is evidenced by the theory's success. In this view, a wave-function represents a physical structure that either exists independently of the more familiar physical systems to which NQMCs pertain or else grounds their existence and properties. From this realist perspective, it may seem natural to label as instrumentalist any approach opposed to that account of the quantum state. But a pragmatist should concede the reality of the quantum state—its existence follows trivially from the truth of quantum claims ascribing quantum states to systems. What he should deny is that quantum state ascriptions are true independently of or prior to the true NQMCs that (in his view) ground them. A more radical pragmatist would reject the representationalist presupposition of this realist/instrumentalist dilemma—the assumption that mere representation is both a (key) function of a novel element of theoretical structure and figures centrally in an account of its content. The truth of a quantum state ascription trivially implies that a wave-function represents *something*, much as the truth of '1+1=2' implies that '1' represents the number 1. By eschewing a more substantial notion of representation this more radical pragmatist could seek to undermine the view that representation of a tolerably insubstantial sort could either be a non-perspectival function of an element of theoretical structure or usefully appealed to in an account of its content.

Not only does quantum theory inform us about the unobserved: it helps agents improve their beliefs about microscopic phenomena that are often considered unobservable. Many NQMCs are about properties of systems that are unobservable by unaided human senses, including at least three of the examples of NQMCs given earlier. This is to be expected, since quantum theory was initially developed as a theory of the microworld, where classical physics was first seen to break down. For the pragmatist, unlike the constructive empiricist or traditional instrumentalist, the observable/unobservable distinction is of no special semantic, epistemic or methodological significance. The use of quantum theory to adjust credences in NQMCs about microscopic phenomena is not only compatible with the present pragmatist view, but plays an important role in helping us explain regularities they exhibit, as we shall soon see.

## 5. How these functions contribute to the explanatory task

The first step in using quantum theory to help explain a regularity is to say what the regularity is. This may be done by an explanandum statement in ordinary or scientific language, perhaps employing some mathematical representation of the regularity to be explained. Before the resources of quantum theory can be brought to bear, it is typically then necessary to *re*present the regularity as involving quantum systems of a certain type or types, and properties of those systems and their environment.[13] The *explanandum* regularity itself must be expressed in suitable NQMCs or non-quantum claims taken to supervene on them, but the circumstances in which it obtains may be described in other non-quantum terms. Cartwright ([1983]) called this first stage of theory entry giving a prepared description of the explanandum and the conditions under which it holds. But note that at this stage we have not yet entered the domain of *quantum* theory, since the prepared description is not of quantum states and Born probabilities and includes no talk of operators.

      When the mathematical modeling apparatus provided by quantum theory is now deployed, the prepared description of the regularity is *not* shown to be a deterministic or stochastic

consequence of laws or principles of quantum theory, dynamical or otherwise. Quantum theory plays a different explanatory role. This is to show that an agent should expect the regularity to hold under these conditions by applying one or more mathematical models of quantum theory to the prepared description. Quantum theory tells an agent what to expect by guiding her credences in NQMCs that have now been taken to represent the *explanandum* regularity. At the same time it tells her what she should have expected in a variety of alternative conditions. It also helps her to see what the regularity depends on and how. It is by exhibiting such dependence that the explanation goes beyond showing *that* the *explanandum* regularity was to be expected by summarizing the physical grounds for this expectation, and so saying *why* it was to be expected.

    The Born Rule plays a key role here: it figures, explicitly or implicitly, in all explanatory applications of quantum theory. Quantum theory contributes to our explanatory projects by providing us with a general set of techniques for calculating Born probabilities that tell us what we should expect, in familiar as well as unfamiliar situations. These include general techniques for assigning the quantum state to a system that is appropriate for an agent in the relevant situation, since the Born Rule can play its role only when supplied with a quantum state. So saying how we use quantum theory to explain involves describing these latter techniques.

    In the foundational literature on quantum theory, this is sometimes known as 'the preparation problem', and seen as complementary to its better known companion, the measurement problem. Suppose one were to follow von Neumann ([1932]) by assuming that measurement projects the quantum state of the measured system onto an eigenstate of the measured observable with eigenvalue equal to the value obtained in the measurement. Then one way to prepare a system in such an eigenstate would be to perform an appropriate measurement and select a system just in

case one obtained the corresponding eigenvalue. But quantum states are often ascribed to systems subjected to no such measurement, and (more importantly) the present interpretation denies that measurement plays any special dynamical role in quantum theory. So what does justify an agent in ascribing a particular quantum state to a system on the present interpretation, according to which unitary evolution would typically rapidly entangle any pure quantum state of a system with the state of its environment?

The general form of the answer is clear, even though details will vary from case to case. A user of quantum theory begins with some prepared description or representation of the physical circumstances within which the quantum state of some system (selected or abstracted by the user) is required. This description or representation will be given in terms of NQMCs and/or other non-quantum claims. This will include a description or representation both of a system and of its environment: for example, a collection of 1 million atoms of rubidium held in a magnetic trap of specified potential at temperature $T$; a pair of photons emerging at particular angles from a suitably prepared potassium niobate crystal on which a laser pulse of specified wavelength, power and duration is incident; a collection of neutrons at the core of a neutron star at specified temperature and pressure; two beams of 3.5 TeV protons of specified intensity and cross section colliding head on in a particular space-time region, in a vacuum, subject to a specified electromagnetic field. This "pre-quantum" description is used to provide premises for a defeasible material inference to a conclusion assigning a particular quantum state to the selected system(s).

To justify the inference, a variety of considerations may be brought forward. These may include arguments that would warrant assigning a particular Hamiltonian operator in a dynamical equation for the system's quantum state, based on assumptions about the interactions to which it is subject, as well as arguments taken to justify imposition of boundary conditions on solutions to that equation. This is where the Schrödinger equation and its relativistic generalizations come in. Often the prepared description contains enough information about the situation in which the explanandum regularity holds to permit one to write down an equation from whose solutions may be correctly inferred the needed quantum states. There are general, though not algorithmic, methods of preparing a description in terms of NQMCs and other non-quantum claims so as to write down Hamiltonian operators and to pick out relevant solutions to the Schrödinger equation.

Such generality enhances the explanatory power we derive from quantum theory by unifying our understanding of all the phenomena to which these methods may be applied. Even simply writing down a plausible-looking wave-function rather than deriving it as a solution to a Schrödinger equation with chosen Hamiltonian can unify one's understanding of a variety of phenomena this helps one to explain: a single wave-function can be used in the Born Rule to calculate probabilities for a wide variety of NQMCs licensed in different circumstances. In favorable conditions, one can derive expectations from some of these probabilities and perform experiments or observations to determine whether these are borne out: if they are not, then this may be judged to defeat the inference to the quantum state that had seemed so plausible.

Note that assignment even of a pure quantum state to a system $S$ in order to guide expectations concerning NQMCs concerning that system is not inconsistent with assignment of an entangled state to a supersystem of $S$ to guide expectations concerning correlations between NQMCs involving $S$. In particular, assignment of a pure quantum state to $S$ may be entirely warranted *for certain purposes* even when $S$ is not isolated but taken to be entangled with another system and/or with its environment. For example, one can justifiably assign neutrons in the "up" beam after passing through an interferometer a spin-up eigenstate, even when the entanglement between their spin and spatial states would have been manifested had the "up" and "down" beams been recombined.

Irrespective of the epistemic procedures used to assign quantum states, whatever non-quantum claims *in fact* ground the correct quantum state assignment thereby serve to specify what any consequent regularities physically depend on. This dependence is both indirect and relative. It is indirect because the connection between these non-quantum claims and an *explanandum* regularity is made through the quantum state they ground and the consequent Born Rule probabilities or their equivalent (e.g. an assignment of expectation values to magnitudes). It is relative to the physical situation of the (actual or hypothetical) agent making the assignment. But despite these two key differences, the use of quantum theory to explain physical phenomena exhibits the physical dependence of those phenomena on conditions described in non-quantum claims. In this respect it does parallel the explanatory application of classical physics.

It is important to understand how the dependence on physical conditions described by true non-quantum claims is mediated by a quantum state in explaining a regularity. These conditions ground the quantum state assignment: they do not cause a system to be in that quantum state. A statement assigning a quantum state to a system at a time is not a description of an event occurring or a physical state present at that time. The statement does not describe an effect of the conditions that ground it but rather supervenes on them.

It is widely acknowledged that one cannot explain a phenomenon merely by showing that it was to be expected in the circumstances. To repeat a hackneyed counterexample, the falling barometer does not explain the coming storm even though alluding to it gives one reason to expect a storm in the circumstances. As a joint effect of a common cause, a symptom does not explain its other independent effects. But a system's being in a quantum state at a time is not a symptom of causes specified by its grounding conditions since it is not an event distinct from those conditions. There is a consequent temptation to say that it is not the quantum state but the conditions that ground it that *really* explain the regularity the quantum state advises one to expect. Here are two reasons not to yield to that temptation.

It is often easier to demonstrate the correctness of a quantum state ascription than to make explicit the non-quantum claims that ground it.[14] The subsequent explanatory application of that state then proceeds independently of the physical grounds of its ascription. One could choose to regard an explanation solely in terms of a quantum state as what Hempel called an explanation sketch: the explanation of which it is a sketch would then go on to specify the non-quantum claims that ground the ascription of that quantum state. But anyone making this choice would have to admit that we have yet to explain many or most of the phenomena we consider successfully explained using quantum theory just by reference to the quantum state.

There is a more important reason to accept an explanation in terms of a quantum state as satisfactory in its own right. Depending on the circumstances, one of a variety of different sets of non-quantum conditions may ground assignment of a given quantum state. We may already know a large number of different ways of preparing that quantum state while confidently expecting to find new ones. The supervenience base for a quantum state ascription is typically large, diverse and open-ended. An explanation in terms of a quantum state acquires unifying power by abstracting from the details of the particular physical conditions in which that state is grounded. Here is a partial analog. An explanation of the behavior of a circuit element that assigns it a resistance and applies Ohm's law is valued for its generality, whether or not one can supplement it by an account of the particular micro-structural behavior on which its resistance happens to supervene. The analogy is only partial because the micro-structural behavior that grounds the element's resistance spatiotemporally coincides with it, while what physically grounds a quantum state ascription to a system at a time typically involves other systems at other times and places.

Suppose a regularity is explained by an agent in terms of a quantum state ascription. If the world had differed so that agent would have been right to ascribe a different quantum state or states to a system, then a different regularity would have obtained, or no regularity at all. These counterfactual dependences on quantum state ascriptions enable the agent to answer many *w*-questions about the regularity. Since in the present pragmatist view quantum states (though objective) don't simply represent properties of the physical systems to which they are assigned, this does not yet say what properties of those (or any other) physical systems the regularity physically depends on. But, as we saw, the correct quantum state ascription itself counterfactually depends on the truth of non-quantum claims stating properties of physical systems. So ultimately the regularity depends counterfactually on the truth of whatever non-quantum claims ground the quantum state ascription, and these counterfactual dependences hold answers to many *w*-questions about what a regularity *physically* depends on.

The dependence of a regularity on a quantum state and on non-quantum claims (including those grounding that state) is relative to the (actual or hypothetical) physical situation of an agent ascribing the state. In particular, agents situated in relevantly different space-time locations should (consistently) ascribe different quantum states to the same system(s) and give different accounts of what the regularity depends on. In such a case, they will explain the regularity differently even though the truth of non-quantum claims (including those on which the regularity depends) is in no way relative to the physical situation of any agent.[15] Such relativity of dependence relations to the space-time location of an (actual or hypothetical) agent presents a difficulty to one who wishes to portray the relevant counterfactual relations as *causal*. At the same time it undermines the view that each instance of a regularity always depends only on what (absolutely) temporally preceded it. On the present pragmatist interpretation, an agent using quantum theory to explain a regularity appeals to a web of counterfactual dependencies among other regularities. What events she takes instances of the *explanandum* regularity to depend on is a function of her own (actual or hypothetical) space-time location, since no such event occurs outside her own past light cone. While this makes the resulting explanation relative to the physical situation of an agent for whom it is tailored, it does not make the explanation subjective. Any agent "projecting himself" into the same physical situation should accept the same explanation.

The assertion that all explanations of physical phenomena using quantum theory are "funneled through" the Born Rule or functional equivalent may strike some as implausibly broad. For example, the explanation of quantization was one of quantum theory's earliest and most striking successes: an observable will always be found to have a value lying in the spectrum of the self-adjoint operator to which it corresponds, and this spectrum often has a pure discrete part. This may seem to be independent of the Born Rule. Redhead ([1987]) gave the following argument for its independence.

> The numbers generated by the quantization algorithm are usually identified with those which turn up in some state with non-vanishing probability according to the statistical algorithm. Or, to put it another way, if the probability of a certain measurement result is always zero, this value cannot be the result of the measurement. But that is just wrong.

Zero probability is quite different from impossibility. (pp. 5-6)

Certainly zero probability does not imply impossibility: if it did, it would be impossible to ionize a hydrogen atom, in so far as every NQMC assigning a precise positive energy to a hydrogen atom in a continuum state has zero probability. But one can still use the Born Rule to explain quantization of the energy of a hydrogen atom by applying it to an arbitrary superposition of bound-state eigenfunctions to show that one should never expect the energy of a bound hydrogen atom to be anything other than a (negative) eigenvalue of the Hamiltonian operator. And one can explain the possibility of ionizing a hydrogen atom by noting that while one should indeed assign zero credence to the atom's having a precise positive energy in any conditions, this is consistent with having non-zero credence for its sometimes having some energy lying within an interval of positive real values. The use of quantum theory to account for quantization of the values of magnitudes involves just a special application of the Born Rule.

A careful examination of other explanatory applications of quantum theory will often reveal a transition from a quantum claim about the expectation value of a magnitude to a corresponding NQMC about the value of that magnitude. Such a move seems most appropriate when the Born probability distribution for that magnitude is sharply peaked about this expectation value, and especially in an eigenstate of the corresponding operator. One such case identifies the momentum of an incoming or outgoing particle with the eigenvalue of a momentum eigenstate. Another example occurs in Ehrenfest's theorem, which purports to derive Newton's second law as a limiting case of the Schrödinger equation for a potential $\phi$ in the form $d\langle\boldsymbol{p}\rangle/dt = \langle-\nabla\phi\rangle$. Such moves are not unwarranted provided one recognizes that quantum theory itself has no implications for the value of any magnitude, but merely offers sound advice as to what one should expect it to be. But acknowledging just this conceptual distinction between quantum and non-quantum claims is vital if one wishes to understand how quantum theory is used, and so how it should be understood.

As the previous section made clear, the Born Rule may be legitimately applied to determine probabilities only of those NQMCs that are suitably licensed in the conditions in which the relevant systems are present. When the Born Rule is used in explaining a regularity, it is the prepared description of the explanandum that specifies the relevant systems and surrounding conditions. Quantum theory furnishes no precise rule specifying exactly when this description legitimizes application of the Born Rule to NQMCs expressing the regularity to be explained. Knowing when it is legitimate to apply the Born Rule here is a learned skill on a par with those needed at an earlier stage in preparing the description of the explanandum regularity and assigning a quantum state to a selected system. One who lacks such skills cannot be said to know quantum theory, no matter how effectively he can write down and solve the Schrödinger equation and calculate Born probabilities from its solutions.

But the theory of decoherence can help to justify a decision to apply the Born Rule under conditions specified by the prepared description of an explanandum regularity. For one can appeal to that description in modeling the effect of environmental interactions on the target systems the description takes to manifest the regularity, as a process by which the coherence of an associated quantum state is rapidly delocalized into the environment. If such decoherence selects a robust set of approximately orthogonal quantum states, each correlated with a particular quantum state (or states) of the environment (or its subsystems), then expectations based on application of the Born Rule to the original quantum state associated with the target systems will be reliably borne out.

Note the role played here by a pragmatist inferentialist account of the content of NQMCs. Whereas in the Newtonian explanation of Kepler's laws the demonstration, within an idealized model, of mathematical representatives of these laws was deductively valid, no deductively valid inference within a quantum model has any NQMC as its conclusion. One cannot even validly infer the objective chance of every NQMC about a system within a quantum model. Rather, the inference to Born probabilities of *selected* NQMCs is justified (given the right kind of decoherence) as a *material* inference that makes its contribution to the content of these NQMCs. And though they are objective, these Born probabilities do not describe or represent physical properties of any system(s), but function solely as authoritative guides to the formation of credences by any appropriately situated agent.

Note also how this pragmatist inferentialist account of the content of such selected NQMCs shows how an explanatory application of quantum theory subtly adjusts the description of the explanandum. What quantum theory helps explain is a claim that is expressed in exactly the language initially used in the prepared description of the explanandum regularity. So the explanation does not correct the initial statement by taking it merely to approximate a statement with which it is strictly inconsistent (as in the classical Newtonian explanation of Kepler's laws). What is corrected is rather the inferential power inherent in that statement. One who accepts the quantum explanation of this regularity acknowledges limitations on what one is entitled to infer from the statement. These limitations will have no practical consequences, since they concern hypothetical situations that are far beyond the powers of any agent to realize (essentially, realizing them would involve reversing massive environmental decoherence—see section 5.2 of (Healey [2012a])). But by helping explain a regularity quantum theory lets us see why it is not quite the statement about beables—'things which can be described in "classical terms", because they are there' (Bell [2004], p. 50)—that it seemed to be. This point deserves a more thorough treatment elsewhere. Nothing in the rest of this paper depends on it.

## 6. Example 1: Single particle interference

While we can use classical physics to explain many interference phenomena involving water, sound, light or other electromagnetic waves, there is a wide range of interference phenomena we can only explain with the help of quantum theory. Some of these are displayed by quantum liquids—many-particle systems in whose behavior not only the effects of quantum mechanics, but also those of quantum statistics, are important. In section 8 I show in outline how quantum theory helps us explain one such phenomenon—interference between separately prepared Bose-Einstein condensates. The present section is concerned with the more familiar phenomenon of single-particle interference. It focuses on two scenarios, one involving relatively massive particles (fullerenes) the other involving single optical photons. I suppress many fascinating details in order to bring out what I take to be the basic form of the explanation in each case, to make clear how quantum theory contributes to the explanation of regularities these scenarios exhibit.

Qualitatively, interference occurs when the result of a process cannot be understood as brought about by a set of what appear to be mutually exclusive and jointly exhaustive alternative subprocesses. The paradigm case is two-slit interference, in which the pattern of fringes on a detection screen placed after a barrier in which are cut two closely spaced, narrow, parallel slits is quite different from the sum of the two one-slit patterns. The difference could be explained quantitatively as well as qualitatively if physical waves of fixed wavelength were incident on the slits, in which case the wave amplitude at the screen would be the vector sum of the amplitudes of two parts of a wave, one part passing through each slit, as long as the intensity of the detected wave is given by the squared norm of its amplitude. But such an explanation is difficult to reconcile with experiments in which single particles are incident on the slits one at a time and then individually detected, and in any case cannot be provided by quantum theory if the wave-function does not represent any physical system or its properties. I shall show how quantum theory helps us to explain interference phenomena manifested in several experiments involving single particles that are incident on more than two slits.

Groups working with Markus Arndt have performed a number of important multiple slit single particle interference experiments with large molecules. In one recent experiment (see diagram 1), Juffman *et al.* ([2009]) prepared a beam of $C_{60}$ molecules with well-defined velocity, passed them through two gratings of a Talbot-Lau interferometer in a high vacuum, and collected them on a carefully prepared silicon surface placed at the Talbot distance. They then moved the silicon about a meter into a second high vacuum chamber and scanned the surface with a scanning tunneling electron microscope (STM) capable of imaging individual atoms on the surface of the silicon. After running the microscope over a square area of approximately $2\mu m^2$ they were able to produce an image of some one to two thousand $C_{60}$ molecules forming an interference pattern. They reported that the surface binding of the fullerenes was so strong that they could not observe any clustering, even over two weeks. Clearly they felt no compunction in attributing very well defined, stable, positions to the molecules on the silicon surface, and even recommended developing this experiment into a technique for controlled deposition for nano-technological applications.

Quantum theory helps us explain the pattern in which the $C_{60}$ molecules were deposited on the silicon surface in this experiment. In rough outline, here is the structure of the explanation. The oven, velocity selector and experimental geometry are taken to warrant assignment to the incident $C_{60}$ molecules a particular initial center of mass wave-function (as a first approximation, a plane wave of corresponding momentum). The geometry of the interferometer supplies boundary conditions on this wave-function as it evolves through the apparatus which enable one to use the Schrödinger equation to calculate its values at the silicon surface. Applying the Born Rule to this wave-function at the silicon surface yields a probability density over NQMCs, each attributing a different position on the surface to a $C_{60}$ molecule. This probability distribution displays characteristic interference fringes of calculable relative magnitudes whose maxima and minima are spatially separated by calculable amounts. An agent who accepts quantum theory should therefore confidently expect that when a large enough number of $C_{60}$ molecules have been deposited on the silicon surface the relative frequencies of molecules in each region of the screen will be very close to the probabilities as calculated from the Born probability density given by the wave-function at the screen (up to experimental error). This explains the regular pattern of $C_{60}$ molecules deposited on the silicon surface—a pattern that is revealed in the subsequent image produced by the STM.

By applying quantum theory here we are able to see what the regularity manifested by the pattern depends on, and to answer many *w*-questions about it. Adjusting the *z*-position of the grating $G_2$ or of the silicon surface would be expected to produce variations in the pattern whose exact character could be calculated by applying the Born Rule to the modified quantum state associated with the different geometry. If the quantum state of the fullerenes had been a mixture rather than a superposition of pure states, each corresponding to passage through a single slit in $G_2$, there would have been no interference pattern. That would have been the case, for example, if the experimental conditions had been different in certain ways. These counterfactual variations in the set-up would also be described in non-quantum claims, such as the claim that gas pressure in the interferometer was raised significantly higher than $10^{-10}$ mbar, or that $G_2$ was subjected to intense illumination during the experiment. The pattern depends also on the mass of the $C_{60}$ molecules: by applying quantum theory here we can see what to expect if we were to replace them with $C_{70}$ molecules, or with a mixture of $C_{60}$ and $C_{70}$ molecules.

While all the dependences in this example considered so far might be taken as causal, other dependences are harder to fit into that mould. The interference pattern here might be expected to differ if the Schrödinger equation were non-linear, or if it were modified by some stochastic collapse term. One can accommodate such a possibility while continuing to deny that the quantum state represents a quantum system or its properties, assigning it the same prescriptive role in advising an agent on what content and credence to give canonical NQMCs. But it is hard to make sense of the idea of intervening to change how this prescription depends on conditions described in non-quantum claims.

Notice that while the Born Rule played a critical role in this explanation, there was no mention of measurement or observation, at least not until the (assumed) final stage at which someone looked at the STM screen on which the image of the deposited molecules was displayed. The explanandum is not that person's subjective visual experience, but the objective pattern in which the $C_{60}$ molecules were deposited on the silicon surface. Though the deposition pattern is clearly unobservable by unaided human senses, any agent aware that a reliably working STM has produced this image is both licensed and warranted in claiming it is an image of an objective pattern formed in the experiment. The license comes from the massive environmental delocalization of the molecules' center of mass position wave-function produced by binding to the silicon surface. The warrant comes from the (assumed) fact that the STM is working well and has been reliably operated.

In other fullerene experiments the interference pattern cannot be explained in quite the same way, since the detectors function not by trapping the fullerenes but by ionizing them and detecting them one by one as each induces the emission of electrons on striking a cathode (see diagram 2). A narrowly focused laser beam ionizes fullerenes only at its "waist" or narrowest point, and the intensity of amplified current is measured as the beam tracks across a detection plane. The laser sets up a strong correlation between the internal energy of a fullerene and its center of mass position in that plane. A fullerene is detected if and only if its internal energy was great enough to ionize it. So detection of a fullerene warrants one in making two NQMCs:

> The internal energy of the fullerene was greater than the ionization energy right after the detection plane.
>
> The position in the detection plane of the fullerene lay in a region very close to the waist of the laser beam.

We have an extensive license to entertain each of these NQMCs given the nature of the massive decoherence of the fullerene wave-function induced by the interaction with the laser beam. The interference pattern recorded by the detector is explained by showing how it is just what one should expect if one applies the Born Rule to NQMCs of the second sort for varying horizontal positions in the detection plane, given the fullerene wave-function there: the Rule also enables one to say in non-quantum terms how the pattern would have differed under alternative physical conditions, including conditions that would have grounded a different quantum state ascription.

A third recent experiment (see diagram 3) has been taken to confirm an important general regularity. In the interference pattern resulting from more than two paths, the interference terms are the sum of the interference terms in the patterns resulting from these paths taken two at a time. The Born Rule provides an explanation of this general regularity: here is a simple instance based on the Born probabilities for the case of three path interference. Label the paths $A,B,C$, the probabilities $P$ (with path subscript) and the interference terms $I$ (with path subscript). Then

$P_A = <A \mid A> = ||A||^2$, etc.

$P_{AB} = <A+B \mid A+B> = ||A||^2+||B||^2+I_{AB}$, etc.

$P_{ABC} = <A+B+C \mid A+B+C> = ||A||^2+||B||^2+||C||^2+I_{AB}+I_{BC}+I_{CA}$

Hence (*I*)  $I_{ABC} = I_{AB}+I_{BC}+I_{CA}$.[16]

Regarding this equation as a claim about statistical distributions, Sinha *et al*. ([2010]) matched interference patterns against an instance of (*I*) in a variety of experiments involving the interference of light at up to three slits. One of these involved single photons, detected in coincidence with a second "herald" photon from an entangled pair by avalanche photodiodes $D_1$, $D_2$. The experimental interference pattern is generated by moving a multimode optical fiber uniformly across a plane intercepting light from the slits and counting the relative number of photons detected in each small region. The recorded statistics conform well to those expected on the basis of three-path instance (*I*) of the general regularity, when one takes this application of the Born Rule to yield probabilities for claims of the form *R*:

(R)     The transverse position of the photon is detected between $r$ and $r+\Delta r$.

But such a claim is not even an NQMC: what licenses one to apply the Born Rule to it here?

This question raises important matters of principle, since talk of photons must somehow be grounded in the quantum theory of the electromagnetic field. Now there is no well-behaved position operator in relativistic quantum theories, and it is difficult if not impossible to understand a relativistic quantum field theory as describing localized particles such as photons.[17] Of course this is to be expected on the present pragmatist interpretation, which denies that it is the function of a quantum theory to describe any physical system. But acknowledging this does not absolve the interpretation of the need to answer the question at issue: How is it that use of the quantum formalism licenses claims about photons, and specifically this claim about a photon's position?

A claim about the position at which a photon is detected may be licensed sufficiently to permit application of the Born Rule by enough decoherence of the right kind in the quantum state of the electromagnetic field. In the present situation, no such decoherence occurs before the field interacts through the photoelectric effect with systems composing the avalanche photodiodes. Such interaction *directly* licenses an NQMC about the energy of an electron ejected through the photoelectric effect in the photodiode $D_2$. But it may be taken thereby *indirectly* to license a claim of the form $R$ in the interception plane across which the multimode optical fiber is tracked. For the "backtracking" inference from ejected electron at photodiode $D_2$ to photon detected at or near position $r$ in the interception plane is warranted here by the assumption that the multimode fiber provided the only available channel through which light could propagate. So it is legitimate to apply the Born Rule to claims of the form $R$ concerning the position of photons in the interception plane, and to take such applications to explain why the actual interference patterns in the interception plane for the 2- and 3-slit configurations (as manifested by counting statistics) conformed to the general regularity for 2- and 3-slit interference patterns that is to be expected on the basis of the probabilistic generalization (*I*).

This third experiment displays regularities at two levels. Each sub-experiment with a particular slit configuration manifests a regular interference pattern that depends on that configuration as well as other fixed elements of the experimental set-up. In addition, the results of the sub-experiments instantiate a statistical regularity corresponding to (*I*). This higher-level regularity is to be expected on the basis of the Born Rule, together with the assumption that the fixed elements of the experimental set-up ground the same initial quantum state in each sub-experiment.

Assuming the Born Rule is a law of nature governing physical systems, one could take the higher-level regularity to nomically depend on the Born Rule. After rejecting this assumption the pragmatist interpretation of (Healey [2012a]) cannot say what this higher-level regularity physically depends on. But there is no need to do so. The Born Rule of quantum theory helps to explain its varied instances, each with its own physical dependences, in basically the same way, and so exhibits a pattern to which they all conform. Here quantum theory helps us to explain not by showing what the higher-level regularity physically depends on, but by exhibiting the pattern to which all of its instances conform as itself a fitting into the still broader pattern of phenomena that quantum theory helps us to explain through use of the Born Rule. Here we see a clear example of how the unifying power of quantum theory contributes to its explanatory credentials.

**7. Example 2: Explanation of the stability of matter**

One of the greatest triumphs of quantum theory is the explanation of the stability of ordinary atomic matter. Its inability to account for this most familiar of regularities was perhaps the most obvious failing of classical physics. As a recent text makes clear (Lieb and Seiringer [2010]), more than one question is raised by the stability of ordinary matter, and fully satisfactory answers to some of these questions have been arrived at using quantum theory only relatively recently.[18] I shall focus on just two such questions, of which this is the simplest: Why is the atom stable against collapse of electrons into the nucleus, given the attractive Coulomb interaction between atomic electrons and nuclear protons? This questions stability of the first kind. Classical physics predicted just such collapse, with an accelerating electron rapidly radiating away an unbounded amount of potential energy as it spiraled into the nucleus.

Attempts to use quantum theory to answer this question typically proceed by offering a proof that the expectation value $<E>$ of the internal energy of an atom has a lower bound. In the simplest treatment, this expectation value is calculated by applying the Born Rule to an arbitrary wave-function that is a solution to the Schrödinger equation for the nuclear protons and atomic electrons subject only to their mutual Coulomb interactions. For the case of the hydrogen atom, the lower bound of $-13.6$ eV is attained by the ground state wave-function. But why does the existence of such a lower bound establish the stability of the atom?

Quantum theory helps one to explain atomic stability by showing why one should never expect the internal energy of an atom to be less than the lower bound $E_{min}$ of $<E>$. An expectation that the internal energy of an atom is less than $E_{min}$ is warranted only if and when an NQMC of the form

(E)     The internal energy of this atom is less than $E_{min}$

is both adequately licensed and warranted: but that is never the case. An NQMC attributing an internal energy to an atom is sufficiently licensed to permit application of the Born Rule only when the atomic wave-function has undergone enough and the right kind of environmental decoherence. But application of the Born Rule in such circumstances will always assign zero probability to ($E$), in which case ($E$) is not warranted. Therefore one who accepts quantum theory should never expect an atom to collapse under the influence of the Coulomb interaction among its electrons and protons. This explains the stability of atoms, since it also indicates what this stability depends on.

The existence of a lower bound for $<E>$ depends on mathematical properties of the Schrödinger equation and also on the form of the atomic Hamiltonian operator that appears in that equation. One can appeal to such mathematical properties in asking and answering a variety of *w*-questions by appeal to the structure of a quantum-theoretic model of the atom. For example, had the two-particle electrostatic contribution to the potential energy term in the Hamiltonian varied as the inverse square of the distance between two oppositely charged particles, the hydrogen atom (for example) would have been unstable. But this is not a way of eliciting *physical* (rather than mathematical) dependence relations if a quantum-theoretic model does not represent the physical properties of an atom. What the stability of the atom *physically* depends on is its composition, as described in non-quantum terms. Part of what is involved in accepting quantum theory is accepting the appropriateness of modeling a system of given composition in a particular way: The quantum state of any isolated system composed of nucleons and electrons should (as a first approximation) be taken to be a solution to the Schrödinger equation for the nuclear protons and atomic electrons with Hamiltonian equal to the sum of the individual kinetic energy operators and potential equal to the sum of the pairwise electrostatic potential energies. Commitment to a particular way of modeling a system described in non-quantum terms is revocable in a way that commitment to the Born Rule and linearity of the Schrödinger equation is not. But in neither case does one extend one's *representational* commitments by accepting this aspect of quantum theory.

A second question concerns what Lieb and Seiringer call stability of the second kind: Why is ordinary matter stable in the sense that the internal energy of $2A$ atoms is twice that of $A$ atoms? If the internal energy of ordinary matter were to decrease faster than linearly with respect to the number of atoms one could release energy by pouring half a glass of water into a half-full glass. The first good explanation of why ordinary matter manifests stability of the second kind was not given until the work of Dyson and Lenard in 1967 (Lieb and Seiringer [2010]). Besides using quantum theory, they had to appeal to the Pauli exclusion principle in order to show that the lower bound on the expectation value of the internal energy of a neutral collection of $N$ electrons and $M$ nuclei was not only finite but also increased linearly with the number of electrons and nuclei.

How do quantum theory and the Pauli principle contribute to this explanation? In outline, the explanation parallels the explanation of stability of the first kind. One applies quantum theory to prove an inequality of the form

(*LIN*)   $E(M+N)_{min} \geq -C(M+N)$,

where $C$ is a positive constant and $E(M+N)_{min}$ is a lower bound on the expectation value $<E_{M+N}>$ of the internal energy of a neutral collection of $N$ electrons and $M$ nuclei.[19] One then reasons that an expectation that the internal energy of such a collection is less than $E_{min}$ is warranted only if and when an NQMC of the form

($E_{M+N}$) The internal energy of $N$ electrons and $M$ nuclei is less than $E(M+N)_{min}$

is both adequately licensed and warranted: but that is never the case. But while the key inequality (*LIN*) holds for electronic wave-functions that are totally antisymmetric under exchange it *fails* for totally symmetric wave-functions. Hence the explanation that ordinary matter displays stability of the second kind depends on the Pauli exclusion principle. Bosonic matter would not possess this kind of stability.

Such appeals to the Pauli principle are common in explanatory uses of quantum theory, perhaps the most famous of which is the application of quantum theory to explain basic aspects of the structure of the periodic table of chemical elements. They prompt the following objection to the present pragmatist view of quantum theory. Since the claim that electrons are fermions is fundamental to quantum theory, quantum theory does indeed expand our ability to represent physical systems, namely electrons. Before quantum theory, we had no such way of representing electrons.

There is a characteristically pragmatist reply to this objection. One who accepts quantum theory will certainly take the statement that electrons are fermions to be objectively true, and since it is about physical systems, namely electrons, it is not simply wrong to take it to offer a new description of them. But the statement functions quite differently from NQMCs about electrons. Its role is to constrain the assignment of quantum states to collections of electrons, or in quantum field theory to the electron field. In this role it acts as a "gate-keeper", overseeing a user's entry into the non-representational mathematics of quantum theory. The Born Rule marks the exit to this mathematics—the point where the user receives quantum-theoretical advice on what descriptive claims it is appropriate to formulate, what he is licensed to infer from them, and what credence he should attach to them. It is such descriptive claims, formulated in NQMCs, that are the target of the mathematical apparatus of quantum theory. Offering good advice to a user on what to make of such claims is the *raison d'être* of that apparatus. Accepting that various statements that occur within the mathematical apparatus of quantum theory should be regarded as true (such as (*I*) and (*LIN*) as well as many other mathematical statements concerning probabilities and quantum states, including the Schrödinger equation) should not be allowed to obscure the vital distinction between the roles played by such statements internal to quantum theory and the descriptive claims about physical systems which are the reason for using the theory.

## 8. Example 3: Bose Condensation

Quantum theory has been successfully applied to account for many otherwise puzzling regularities involving the behavior of bulk matter at low temperatures. These include resistanceless flow of liquid $^4$helium at low temperatures, the expulsion of magnetic flux by tin as it is cooled below 3°K, and the interference fringes formed when light is shone through two overlapping samples of cold, dilute rubidium gas, each just released from a separate part of a magnetic trap. These and many other regularities in the behavior of superfluids, superconductors and cold, dilute alkali gases are attributed to the formation of a Bose-Einstein condensate (BEC)—a state of matter whose existence Einstein first predicted using quantum theory.

When using quantum theory to explain such regularities it is essential to assign the right kind of wave-function to the system concerned. In order for the Born Rule to yield the right bosonic probabilities this must this be totally symmetric under all pairwise exchange of "particles" (e.g. $^4$helium atoms or Cooper pairs of electrons). But this alone is not enough. To account for a phenomenon involving Bose condensation this wave-function must be one in which "a macroscopic number of particles occupies the same one-particle state". Leggett ([2010], p. 31) puts this last condition in scare quotes prior to offering criteria for being a BEC that make it more precise. I shall give his favored criterion for spinless particles, but first note that I used scare quotes here for a different reason. In classical physics a particle may be said to occupy a state if and only if it possesses the dynamical properties characterizing that state, as spelled out in NQMCs. But to assign a quantum state to a system is not to represent it as having any dynamical properties, at least on the present pragmatist interpretation of quantum theory. An agent assigns an appropriate quantum state to a system simply in order to use quantum theory as a source of good advice on the content and credibility of NQMCs concerning that system. So it is highly misleading to talk of particles occupying quantum states, as if this simply described their intrinsic physical properties.

Leggett defines what it is for a physical system to exhibit simple BEC by reference to the most general quantum position-representation pure-state wave-function for a system of $N$ spinless bosons:

$$\Psi_N(t) = \Psi_s(r_1, r_2, \ldots, r_N : t),$$

where the $s$-subscript indicates that the wave function $\Psi_s$ is symmetric under exchange of any pair of position vectors $r_i, r_j$. The most general such mixed state can then be expressed as a convex combination of such normalized and mutually orthogonal pure states with weights $p_s$. This gives rise to the (un-normalized) single-particle density matrix $\rho(r, r' : t)$ defined by

$$\rho(r, r' : t) = N \sum_s p_s \int dr_2 \, dr_3 \ldots dr_N \, \Psi_s^*(r, r_2, \ldots, r_N : t) \Psi_s(r', r_2, \ldots, r_N : t)$$

which may be written in diagonal form, where the $\chi_i(r\,t)$ are orthonormal eigenfunctions

$$\rho(r, r' : t) = \sum_i n_i(t) \, \chi_i^*(r\,t) \, \chi_i(r't)$$

A system of $N$ spinless bosons is said to exhibit simple BEC at time $t$ if and only if there is exactly one eigenvalue $n_i(t)$ of order $N$ while the remaining $n_j(t)$ ($j \neq i$) are all of order 1.

This criterion for what it is to exhibit simple BEC places three important conditions on a quantum state assignment to a system of $N$ particles. The first condition, that the particles are bosons, should be understood in basically the same way as the condition that electrons are fermions discussed in the previous section. While such statements may appropriately be judged as true, their primary function within quantum theory is to constrain quantum state assignments, and thereby the Born probabilities for NQMCs that follow from them. The same analysis applies also to the second, main condition on the eigenvalues of the single-particle density matrix: while a statement assigning a quantum state to a system may be objectively true or false (relative to a given agent-situation), its function in quantum theory is not to ascribe any property to that system, or relation to that agent-situation. A final condition on this initial definition of BEC is that it applies to spinless particles. The statement that $^4$helium atoms are spinless certainly looks like a descriptive claim about $^4$helium atoms. But once again, the function of a statement attributing a particular spin (or lack of it) to a quantum system is not to ascribe an intrinsic, characteristically quantum, *property* to that system, but to instruct an agent on the appropriate Hilbert space to use when assigning quantum states to systems of that type. In sum, none of the three conditions involved in this criterion for BEC describes or represents any physical property of these systems.

To *apply* a criterion for BEC like Leggett's a user of quantum theory needs to be able to tell when the condition it specifies is satisfied. This involves a defeasible material inference from a suitably prepared description in terms of NQMCs and other non-quantum claims to the assignment of a quantum state satisfying that condition: In the material mode, it involves saying when a system exhibits BEC. For real systems, this can be difficult: in the case of high-temperature superconductors, the required inferences are still in need of clarification.[20] It may be helpful to consider instead the simpler case of an ideal, non-interacting system of bosons in thermal equilibrium with a heat bath at temperature $T$.

At sufficiently low temperature $T$, quantum statistical mechanics assigns this system a mixed quantum state of which the largest component is a product of ground state energy eigenstates of individual bosons. Interaction with the heat bath decoheres the energy states of the system to license NQMCs about its energy. So the Born Rule warrants the claim that a significant fraction of the system is composed of bosons, all with the same ground state energy. Interactions with such a BEC that decohere its quantum state in a position basis (e.g. by shining laser light on it) would license NQMCs about BEC density consistent with the Born probability density corresponding to the single particle quantum ground state.

Bose condensation of a cold, dilute alkaline gas was first achieved in a laboratory only quite recently (Anderson, M.H., *et al.* [1995]). In this system, the bosonic atoms do interact, but otherwise the situation approximates the ideal considered in the previous paragraph. Quantum theory helps us explain interference between separately prepared BEC's of this kind. Such interference was observed by Andrews *et al.* [1997], who collected rubidium atoms on each side of a magnetic trap split into two by a strong laser beam, and cooled them until each of the two collections separately underwent a process of Bose condensation. After each condensate reached thermal equilibrium, the laser was turned off so the condensates could both expand and overlap. Light was passed through the overlapping condensates, some of which was absorbed. The remaining light displayed periodic variations of intensity with position along the axis joining the two parts of the trap. This was taken as evidence of density fluctuations in the merged condensate indicative of interference between the two condensates.

How can quantum theory help to explain the observed interference pattern? Its origin is quite different from the single-particle interference patterns considered in section 6. Any explanation begins by assigning a fully symmetrized wave-function to the separately prepared condensates to ensure satisfaction of the bosonic constraint. While alternative state assignments have been proposed, I follow the analysis of Leggett ([2010]) and others who defend the assignment of a double Fock state with no initial phase relation between the states of the two condensates. The resulting Born probability density for single particles contains no interference terms, which initially seems to render the repeatable observation of interference fringes inexplicable. But the *joint* probability density for multiple particles displays correlations that become more pronounced the larger the number of particles one considers. Now light absorption by a condensate is a process that involves an individual interaction with each of very many atoms. Metaphorically speaking, it therefore probes the strong many-particle correlations in the joint Born probability densities rather than the non-existent interference terms in the single-particle Born probability density. If the experiment were performed very many times under identical conditions one would expect to see no interference after the results of all the experiments were superimposed. In each run of the experiment, one would expect to see an interference pattern, whose overall phase varies randomly from one run to another.[21]

Indeed, the interference patterns produced in this experiment manifest two regularities. On each occasion a pair of condensates is made to overlap, an interference pattern appears with the same well-defined spacing between adjacent fringes; while the exact location of the maxima varies randomly from one occasion to the next. Both these regularities are to be expected, given the ascription of an initial double Fock state to the condensates when the laser is turned off. The fringe spacing depends on the difference between a wavelength associated with the single-particle wave-function of one condensate and that of the other. This in turn depends on the conditions under which the condensates were prepared, including the fact that they were composed of this isotope of rubidium rather than some other alkaline gas isotope, as well as on the nature of the trap. The *absence* of any interference pattern in the *summed* statistics of repeated iterations of the experiment under the same conditions depends on the *absence* of any initial phase relation between the Fock state wave-functions of the two condensates. This is a straightforward consequence of the fact that they were independently prepared in different parts of the trap.

      The NQMCs on which these many-particle joint Born probability densities offer advice concern the axial locations within the condensate at which atoms of rubidium were present when absorbing the incident light. One cannot legitimately infer that the rubidium atoms always have definite locations within the condensate—indeed the only NQMC generally warranted here attributes to each atom a diffuse location throughout the condensate. But the interaction resulting in light absorption induces decoherence of the condensate wave-function sufficient to license application of the Born Rule to much more precisely defined NQMCs concerning atomic positions. This application of the Born Rule to sufficiently many of these claims locating particles within the condensate at the moment when light is incident on it after the original condensates have merged will warrant the firm expectation that the "clumped" positions they then assume will conform to an interference pattern—the "negative" of the pattern recorded by the light that is not absorbed but recorded.

## 9. Conclusion

Among contemporary philosophical accounts of scientific explanation, two approaches have been popular. The first takes scientific explanation to be a matter of exhibiting the causes of a phenomenon. The second views the explanatory project globally, as we unify our knowledge by using a scientific theory again and again to show how what may have appeared to be unrelated phenomena all in fact occur for similar reasons.

Quantum theory has not been kind to causal accounts. Quantum indeterminism seemed to rule out any strict regularity view of causation, and even probabilistic theories of causality have run into severe difficulties in light of Bell's theorem.[22] There are deterministic interpretations of quantum theory, and proposals to reconcile indeterministic causation with Bell's theorem. But apart from their other deficiencies these all subject the concept of causation to severe strain.[23] The account presented here of how we use quantum theory to explain regularities makes no attempt to portray these explanations as causal. But it is consistent with the intuition that such explanations display two kinds of dependence of the regularities on the conditions in which they are manifested.

The regularity depends on the quantum state or states that figure in the explanation: these states are what give one reason to expect the regularity to obtain. This kind of dependence is certainly not causal, since a quantum state does not represent or describe the momentary condition of a physical system to which it is ascribed. But it is still physical, in the sense that the *correct* quantum state ascription supervenes on the (non-quantum) physical conditions that ground it.

The function of the quantum state is to offer advice to a situated agent on the content and credibility of claims about the physical world. The quantum state can serve this epistemic function even if the agent does not know all the physical conditions that ground that state. But the advice is *sound* only if the agent ascribes the correct quantum state, and it is conditions in the physical world as spelled out in non-quantum terms that determine the correct quantum state for her to ascribe in her situation. Non-quantum claims stating the conditions on which a regularity physically depends typically describe systems other than the system to which a quantum state is ascribed in explaining that regularity. None of them describe the quantum state itself. We may aim for a richer account of how quantum theory explains a regularity by making explicit in non-quantum terms the conditions on which that phenomenon physically depends. A more complete explanation of a regularity using quantum theory will show how the regularity counterfactually depends on these conditions themselves. By manipulating these conditions we can sometimes produce an instance of the regularity, as Andrews *et al.* [1997] produced interference between two BECs. In this case it is hard to deny that the experimenters caused the interference. But it is misleading at best to describe as causal the explanation quantum theory helps us to give, either of this interference or of other phenomena including many in the universe we are not able to produce.

When quantum theory helps explain a regularity involving a system, the condition of the system's physical *environment* is important. Environmental decoherence plays a key role in licensing NQMCs that express the regularity to be explained.[24] To finally dispose of the notorious quantum measurement problem more also needs to be said about just what conditions justify the ascription of a particular quantum state. How quantum theory helps explain so-called non-local quantum phenomena involving entangled systems will be the subject of a companion paper that says more about the dependency relations these involve.

Quantum theory helps us explain in ways that clearly unify our knowledge, and thereby contributes to our understanding of the phenomena we use the theory to explain. Despite their important representational differences, models of quantum theory and models of classical physics unify our understanding in very similar ways when deployed in theoretical explanations. In quantum models we don't *represent* physical systems and their behavior by wave-functions, operators, Schrödinger's equation or Born probabilities. But we *use* these and the mathematical framework within which they operate again and again in much the same way when applying the quantum models in which they figure to physical systems in order to see what to expect of them and what this depends on. We don't literally *represent* a half-silvered mirror by a beam-splitter Hamiltonian: instead we come to *mentally associate* half-silvered mirrors, superconducting resonators and suitably tuned partially overlapping laser beams with the same type of term in the Hamiltonian of quantum models that we apply in much the same way to explain their otherwise diverse physical behavior.

Are quantum explanations structural? Yes and no. Elements of a quantum theoretic model neither resemble anything in the physical world, nor are they (even partially) isomorphic to them in any non-trivial sense of that systematically ambiguous term. So I reject any account of structural explanation that appeals to an interesting model-world matching relation.[25] But structural aspects of the mathematical models employed in quantum theory are critical to their explanatory power.[26] By exploring the structure of the abstract mathematical spaces in which quantum states and operators are defined we can come to understand the ways that Born probability assignments to various NQMCs hang together. By itself this explains no physical phenomenon, since Born probabilities are just numbers, absent the physical conditions required for them to offer authoritative advice to an agent on how to apportion credences concerning certain of these NQMCs.[27] But it is by recognizing the commonality of abstract mathematical structure inherent in the family of models explanatorily deployed in a variety of physical contexts that we are able to gain a unified understanding of the diversity of phenomena thereby explained. That is the sense in which quantum explanations are indeed structural, a point that becomes especially clear when one *contrasts* the abstract structures inherent in quantum models with those inherent in models of classical physical theories.

Philosophy Department,
University of Arizona,
213 Social Sciences,
Tucson, AZ 85721-0027, USA.
rhealey@email.arizona.edu

**Acknowledgements**

Embryonic versions of this material were presented in June, 2011: in Brandeis, Massachusetts at a conference honoring Hilary Putnam's 85th birthday and to the Oxford philosophy of physics seminar. I thank those audiences, Paul Teller, Simon Friederich, Terry Horgan, Jenann Ismael, participants in a seminar at Princeton University in Fall 2012, and several referees for this journal for their critical comments. I, at least, think they have resulted in a better paper! I thank Markus Arndt for permission to reproduce Diagram 1. This publication was made possible through the support of a grant from the John Templeton Foundation. The opinions expressed in this publication are those of the author and do not necessarily reflect the views of the John Templeton Foundation.

Diagram 1

Copyright 2009 by the American Physical Society.

**Diagram 2**

Reprinted by permission from Macmillan Publishers Ltd: *Nature*, **401**, p. 680 copyright 1999.

**Diagram 3**

From Sinha, U. *et al.* [2010]. Reprinted with permission from AAAS.

¹ As does Peres ([1993], p. 13), who points out how this differs from an understanding of the theory according to which it can explain some properties of bulk matter whenever these can be derived from those of the microscopic constituents.

² See Deutsch ([1997], [2011]), Saunders *et al.* (*eds.*) ([2009]) and Wallace ([2012]) for examples of the former; and Durr *et al.* ([1992]), Maudlin ([2011]) and Ghirardi *et al.* ([1986]) for examples of the latter. A referee suggested that a proponent of some such realist interpretation may come to think of the pragmatist view of Healey ([2012a]) as an attempt to develop a non-instrumentalist account of quantum theory which proponents of realist interpretations could agree is explanatory in some minimal sense, while continuing in their rival attempts to extract deeper explanations from quantum theory in terms of additional ontological elements (such as quantum states, Born probabilities or particle trajectories). Ecumenism is certainly a worthy aim. But in this case the objections raised against each realist interpretation by proponents of others have so far convinced me that a realist interpretation of quantum theory is simply not to be had. Perhaps some successor of quantum theory will admit a realist interpretation. Meanwhile, I take the philosophical lesson of quantum theory to be that only a pragmatist view enables one to understand the unprecedented explanatory as well as predictive success that makes us call it fundamental.

³ See Hughes ([1989], [2010]), Clifton ([unpublished 1998]), Dorato and Felline ([2011]).

⁴ Strevens ([2008]) similarly lumps together explanations of particular events and general regularities within a causal framework: 'The explanation of a causal generalization [regularity] and the explanation of any instance of the generalization invoke the same causal mechanism' (p. 223).

⁵ There are many targets for non-causal explanations in physics: Why are inertial and gravitational mass equal in Newtonian physics? Why is the speed of light the same in every inertial frame? Why is electric charge quantized? Why are there three generations of quarks and leptons? In such cases, a satisfactory explanation is not expected to point to any possibility of intervention, no matter how recondite. By narrowing his focus to causal explanations, Woodward need not address these cases.

⁶ I will argue elsewhere that since key outcome-outcome counterfactuals that figure in so-called "non-local" correlations should not be understood causally, on the present pragmatist interpretation one can use quantum theory locally to explain these correlations without violating Bell's ([2004]) local causality condition.

⁷ As section 5 explains, according to the pragmatist interpretation of (Healey [2012a]) accepting quantum theory requires reconfiguring material inferential relations among statements, including those representing a regularity we use the theory to explain. This modifies the content of these statements so they are better able to express what needs to be said in a "quantum world". To pursue this point here would prove an unnecessary distraction from the goals of the paper.

⁸ Duhem ([1906], p. 50) notoriously required an explanation to 'strip reality of the appearances covering it like a veil, in order to see the bare reality itself'. Wittgenstein ([1922], 6.371) famously remarked in his *Tractatus* that 'at the basis of the whole modern view of the world lies the illusion that the so-called laws of nature are the explanations of natural phenomena'. But this remark is not easily squared with the quietist philosophy he developed after repudiating his earlier work.

⁹ Cf. Kuhn ([1977], p. 29). Huygens and others rejected the explanatory claims even of Newton's new theory of light and colors, let alone his theory of motion and gravitation.

[10] Note that this is a *negative* characterization of NQMCs: any claim about a magnitude counts as an NQMC provided that it does not concern a quantum state, quantum probability (or expectation value) or other model element introduced by quantum theory (e.g. operators on Hilbert space or elements of some more general quantum algebraic structure). Some NQMCs concern magnitudes of classical physics: with the progress of physics we may confidently expect to be able to make other NQMCs of a kind unknown to classical physics. Healey [2012a] included among NQMCs only those expressible in the canonical form *The value of M on s lies in set* $\Delta$ (*M$\varepsilon\Delta$*). While the Born Rule itself directly assigns probabilities only to such claims, its application often warrants an agent in forming expectations about the values of magnitudes expressible only in other terms, as these examples indicate.

[11] Here I am indebted to Price's subject naturalist approach to truth and representation ([1988], [2011]) which admits a wide range of truth-evaluable discourse while emphasizing the need to understand the diversity of functions such discourse may serve.

[12] Unlike dynamical variables, operators are not even candidates for observation, though the systematic ambiguity encouraged by the term 'observable' tends to hide this fact. Observing statistics for many measurements of dynamical variables permits reasonable though defeasible inferences to claims about Born probabilities and quantum states. But quantum states are neither physical systems nor properties of them, and no propensities underlie Born probabilities.

[13] While quantum *mechanics* may be applied directly to systems such as atoms, electrons and dilute alkaline gases described in NQMCs stating regularities to be explained, systems of quantum *fields* are not themselves described by NQMCs in explanatory applications of quantum field theory

(according to the pragmatist interpretation of (Healey [2012a])).

[14] Parametric down-conversion of laser light in a non-linear crystal has become a standard technique for preparing known entangled quantum states for which it is both difficult and practically unnecessary explicitly to state the exact non-quantum grounding. The Laughlin wave-function proved to be remarkably successful in predicting and explaining features of the quantum Hall effect, although it was proposed without any precise specification of the conditions grounding it.

[15] Such relativity to their space-time locations is what permits agents locally to explain so-called "non-local" correlations, according to the pragmatist interpretation of (Healey [2012a]): see (Healey [forthcoming]).

[16] This deduction of (*I*) from the Born Rule conforms to Hempel's ([1965]) model of what he calls deductive statistical explanation. It uses the mathematical theory of probability to infer a probabilistic generalization from another probabilistic generalization of wider scope. In the pragmatist view of Healey [2012a] this deduction alone offers a mathematical but not a physical explanation, since (*I*) does not express a statistical regularity, but a constraint on a rational agent's credences.

[17] See, for example Malament ([1996]), Clifton and Halvorson ([2001]), Fraser ([2008]), Ruetsche ([2011]).

[18] Indeed, as Lieb ([2007]) explains, many things have yet to be done to understand the stability of matter in the context of quantum electrodynamics. I thank Tony Sudbery for emphasizing the limitations of text-book explanations of the stability of ordinary atomic matter that apply quantum

theory to particulate constituents of atoms while treating their electromagnetic interactions classically.

[19] This discussion is slightly oversimplified. In fact, $C$ depends on the maximal nuclear charge of the $M$ nuclei.

[20] Since the electrons in a superconductor are fermions rather than bosons, the above definition of BEC already needs to be modified to apply to condensing Cooper pairs in a "classical" superconductor, as Leggett ([2006]) does in section 2.4, where he refers to condensed pairs of fermions as exhibiting pseudo-BEC. In the absence of a theory like the BCS theory of "classical" superconductors, the conditions under which condensation occurs in a "non-classical" superconductor may as yet only be partially understood in phenomenological terms.

[21] Healey ([2011]) discusses a closely related example in greater detail.

[22] See, for example, (Healey [2009]). Causal process and conserved quantity accounts of causation such as those of Salmon ([1994]), Dowe ([2000]) also have trouble accommodating quantum explanations.

[23] Bohmian theories are explicitly non-local and violate fundamental Lorentz invariance. GRWP with flash ontology (see Tumulka ([2006]), Maudlin ([2011], chapter 10)) preserves relativistic invariance but is perhaps best seen as an interesting but rather *ad hoc* competitor to a fragment of quantum theory, incorporating (at best) a symmetric causal dependence between isolated point events involving no persisting objects and mediated by no localized causal process.

[24] Healey [2012b] says more about how environmental decoherence licenses NQMCs.

[25] Such as those of Hughes ([1989], [2010]), to whom we owe the present usage of the term 'structural explanation', though the notion of representation did not loom large when he first introduced it, and Dorato and Felline ([2011]).

[26] As Clifton's (unpublished [1998]) lovely example of the hyper-entanglement of the quantum field-theoretic vacuum makes abundantly clear. In fairness, Dorato and Felline ([2011]) are charitably interpreted as making the same point, despite their repeated appeal to representation relations.

[27] This places Clifton's (unpublished [1998]) example of the hyper-entanglement of the quantum field-theoretic vacuum in a somewhat ironic light, since in a physically possible world whose state is the quantum vacuum there are presumably no agents, and (more significantly) no decohering environment to which an agent in *this* world is implicitly appealing in any explanatory application of quantum theory to that possible world. This does not detract from Clifton's main point, which doubtless applies also to less idealized (and therefore more physically relevant) states of a quantum field.

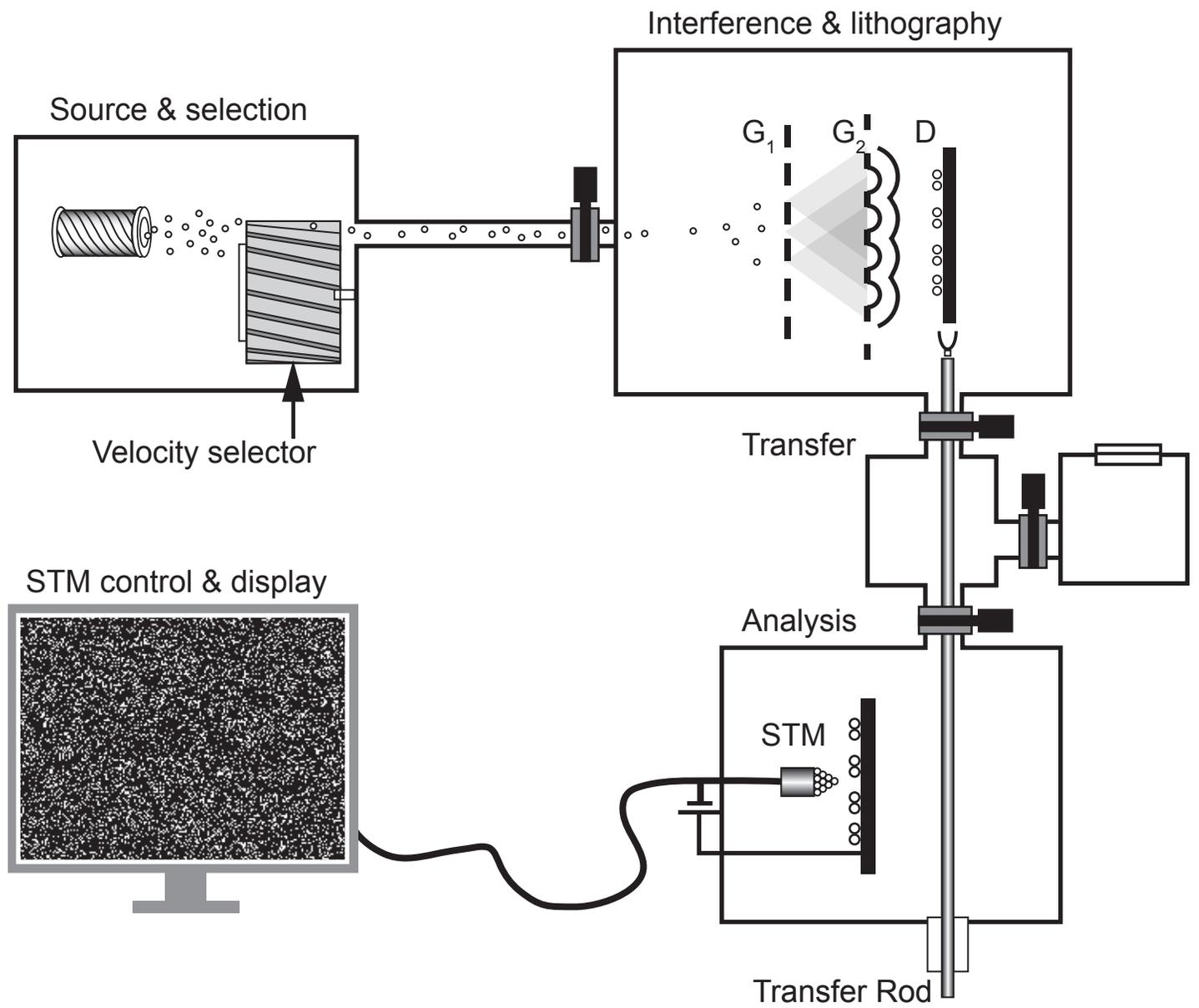

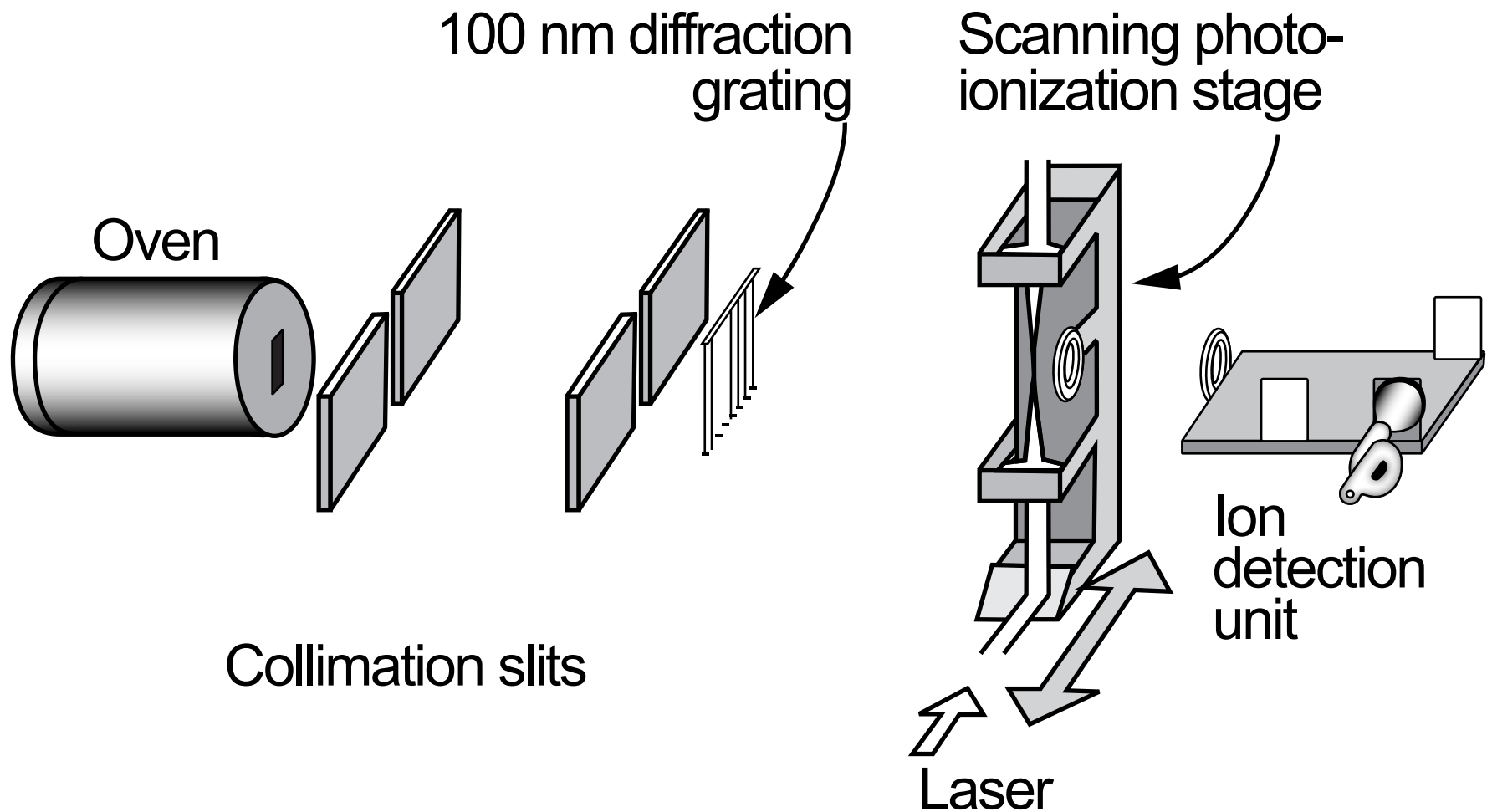

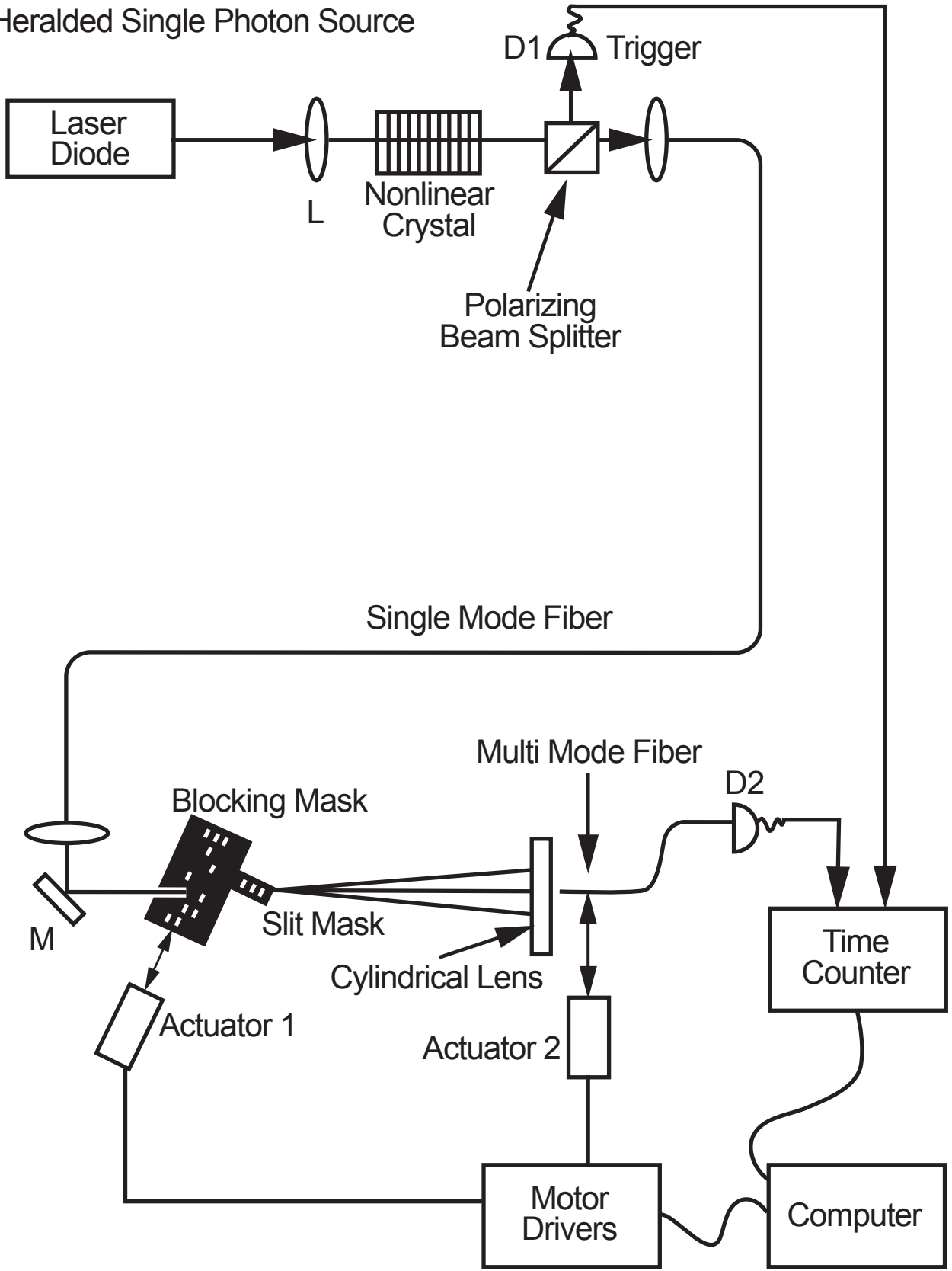